\documentclass[a4paper,10pt]{article}

\usepackage[round,authoryear,sort]{natbib}
\usepackage{graphicx}
\bibpunct{(}{)}{,}{a}{,}{,~}
\newcommand{\forget}[1]{}

\title{Ambiguities in order-theoretic formulations of thermodynamics} 
\author{ Robert Marsland III\\
(Physics of Living Systems Group, 
Massachusetts Institute of Technology),\\  
Harvey R. Brown\\ 
(Faculty of Philosophy, University of Oxford)\\
Giovanni Valente \\
(Department of Philosophy, University of Pittsburgh)}

\date{}

\begin{document} 
\maketitle

\begin{abstract}Since the 1909 work of Carath\'{e}odory, formulations of thermodynamics have gained ground which highlight the role of the the binary relation of adiabatic accessibility between equilibrium states. A feature of Carath\'{e}dory's system is that the version therein of the second law contains an ambiguity about the nature of irreversible adiabatic processes, making it weaker than the traditional Kelvin-Planck statement of the law. This paper attempts first to clarify the nature of this ambiguity, by defining the arrow of time in thermodynamics by way of the Equilibrium Principle (``Minus First Law''). It then argues that the ambiguity reappears in the important 1999 axiomatisation due to Lieb and Yngvason. 
 
\end{abstract}

\section{Introduction}

A number of attempts in recent decades to put equilibrium thermodynamics on a rigorous conceptual and mathematical footing have been been influenced by the 1909 work of Constantin Carath\'{e}odory\footnote{Carath\'{e}odory (1909).}. A prominent example is that due to Elliott Lieb and Jacob Yngvason, in a lengthy paper published in 1999\footnote{Lieb and Yngvason (1999).}. These authors follow Carath\'{e}odory in basing their approach on the notion of adiabatic accessibility, but do without the machinery of differential forms that Carath\'{e}odory had used in his reasoning. A penetrating analysis of the Lieb-Yngvason formulation was published by Jos Uffink in 2001, principally with a view to determining which axioms proposed by these authors were time symmetric and which not.\footnote{See Uffink (2001), section 11. A critique of this analysis and a related one in Brown and Uffink (2001) is found in Henderson (2014).} In the present paper, we are concerned with a related but distinct issue. It is well known that Carath\'{e}odory's formulation contained an ambiguity, or incompleteness, which Carath\'{e}odory himself highlighted, and which is connected with the fact that his postulates lead to a version of the second law that is weaker than the traditional version due to Kelvin and Planck. Some Carath\'{e}odory-inspired formulations of thermodynamics, whether of the formal\footnote{See for example Boyling (1972), pp. 42-43.} or informal variety\footnote{See for example Turner (1962) and Sears (1966).}, add an extra postulate to remove the ambiguity. We argue that this is implicitly what  Lieb and Yngvason do in their approach.

\section{Carath\'{e}odory}

In his seminal 1909 reformulation of thermodynamics, Carath\'{e}dory realised that heat need not be introduced as a primitive notion, and that the theory could be extended  to systems with an arbitrary number of degrees of freedom using generalised coordinates analogous to those employed in mechanics. In doing so, he provided the first satisfactory enunciations of what are now called the zeroth and first laws of thermodynamics.\footnote{See Uffink (2001), section 9.} In particular, by defining an adiabatic enclosure in terms of its capacity to isolate the thermodynamic variables of the system of interest from external disturbances, Carath\'{e}odory was the first to base the first law, and thus the existence of internal energy, on Joule's experiments (under the assumption that Joule's calorimeter was adiabatically isolated). Heat is then defined as the change in internal energy which is not accounted for by the work being done on or by the system, the existence of heat being a consequence of the first law and the conservation of energy.

However, what is of particular relevance for our purposes is that Carath\'{e}odory did to the equilibrium state space something akin to what his ex-teacher Hermann Minkowski had done to space-time a year earlier.
Assuming that the space $\Gamma$ of equilibrium states is an $N$-dimensional differentiable manifold equipped with the usual Euclidean topology, Carath\'{e}odory introduced the relation of adiabatic accessibility between pairs of points, a notion clearly analogous to that of the causal connectibility relation in Minkowski space-time.\footnote{See Uffink (2001). A stronger analogy is with the geometric treatment of the conformal structure of special relativistic space-time found in Robb (1921); see Uffink (2003), p. 144.} He famously postulated that in any neighbourhood of any point $p$ in $\Gamma$, there exists at least one point $p'$ that is not adiabatically accessible from $p$, so that every point in $\Gamma$ is an ``i-point'' in the terminology of Landsberg\footnote{This is the way the postulate is commonly stated in the literature. Carath\'{e}odory's actual axiom was ``In every arbitrarily close neighbourhood of a given initial state there exist states that cannot be approached arbitrarily closely by adiabatic processes." Carath\'{e}odory (1909) p. 236.}. An adiabatic process is one taking place within an adiabatic enclosure.

Carath\'{e}odory's main result concerns ``simple'' systems, whose states can be described by a single thermal coordinate along with an arbitrary number of ``deformation'' coordinates, sometimes called work or configuration coordinates, which depend on the external shape of the system and on any applied fields. This rules out systems comprised of a collection of subsystems adiabatically isolated from each other. Carath\'{e}odory also assumed that simple systems show no internal friction or hysteresis in sufficiently slow (quasi-static) processes. As a result of these and other assumptions, he showed that quasi-static processes involving simple systems can be represented by continuous curves in the state space, where the external work associated with the process can be determined solely by the forces required to maintain equilibrium at all times. (Carath\'{e}odory made a point of \textit{proving} that quasi-static adiabatic processes of a simple system are reversible.) By appealing to a result in the theory of Pfaffian forms, Carath\'{e}odory was further able to show that given the i-point principle, the differential form for heat for quasi-static processes has an integrating factor. In other words, there exist functions $T$ and $S$ on the state space such that the heat form can be expressed as $T\mbox{d}S$, where $\mbox{d}S$ is an exact differential. Further considerations show that $T$ and $S$ are related to the absolute temperature (which depends on empirical temperature as defined by way of the zeroth law\footnote{Turner (1963) has argued that the zeroth law is not strictly necessary in Carath\'{e}odory's treatment.}) and entropy of the system.\footnote{A modern geometric treatment of Carath\'{e}odory's theorem in the context of the Frobenius theorem and nonholonomic constraints is found in Frankel (2004), sections 6.3d-6.3f, in which the energy ambiguity is again resolved by appeal to experience (p. 186).}

\section{The ambiguity}

In section 9 of his 1909 paper, devoted to irreversible processes, Carath\'{e}odory introduced a terse argument related to simple systems that has often been repeated and/or elaborated in the literature.\footnote{See, for example, Landsberg (1990), section 5.2., Turner (1960) and Sears (1963).} The conclusion of the argument is that, given the i-point postulate and certain continuity assumptions\footnote{For a discussion of these assumptions, see Landsberg (1961).}, then for any two points $p$ and $p'$ not connected by a reversible quasi-static path, when $p'$ is adiabatically accessible from $p$, always either $S(p') > S(p)$ or $S(p') < S(p)$. (Quasi-static adiabatic processes involve no change in entropy.) Regarding this ambiguity, Carath\'{e}odory emphasised both that it persists even when the entropy is defined so as to make the absolute temperature positive, and that it can only be resolved by appeal to experiment:
\begin{quote}
Experience (which needs to be ascertained in relation to a single experiment only) then teaches that \textit{entropy can never decrease.}\footnote{Carath\'{e}odory (1909), section 9. The recourse to experiment is further justified in Carath\'{e}odory (1925); a discussion of historical responses to this feature of his theory is found in Uffink (2001).}
\end{quote}

It is important for our purposes to note first that the prior existence of an entropy function is not in fact intrinsic to the argument, or rather that a related ambiguity can be derived in a more general way. The single thermal coordinate for the simple system in question could be chosen instead to be internal energy (whose existence is a consequence of the first postulate in Carath\'{e}odory's paper). In this case the i-point postulate and the same continuity assumptions can be shown to result in the existence of a foliation of $\Gamma$ (subject to a qualification to be clarified below), such that on each hypersurface any continuous curve represents a reversible, quasi-static adiabatic process involving a continuous change in the deformation coordinates. In the case of an \textit{arbitrary} adiabatic process from  $p$ to a distinct state $p'$, these states will generally not lie on the same hypersurface associated with the foliation, but it can be shown from Carath\'{e}odory's postulates that they must all lie on the same side of this surface. In particular, when  $p$ and $p'$ share the same deformation coordinates, $p'$ will either always correspond to greater internal energy than $p$, or always less internal energy, independently of the choice of the initial state $p$. Let us call this \textit{the energy ambiguity} for adiabatic processes. 

A related ambiguity holds when the thermal coordinate is chosen to be temperature (empirical or absolute in Carath\'{e}odory's terms, but assumed to be positive). Indeed the underlying ambiguity in Carath\'{e}odory's formulation of thermodynamics -- prior to the performance of the ``single experiment'' referred to above, and given the positivity of temperature, can also be stated as: Either heat always flows from a hot body to a cold body, or the converse. When considering cyclic processes,  the ambiguity can be expressed in two further ways:
\begin{enumerate}
 \item Either it is always impossible to create a cyclic process that converts heat entirely into work,
or it is always impossible to create a cyclic process that converts work entirely into heat\footnote{See Landsberg (1956) and Dunning-Davies (1965)}; and
\item In relation to a Carnot cycle, any other type of cyclic process either always has lower efficiency, or always greater efficiency\footnote{See Ruark (1925)}. 
\end{enumerate}
Statement 1 is clearly weaker than the traditional Kelvin-Planck form of the second law in thermodynamics; indeed  Carath\'{e}odory's i-point postulate above is easily seen to be a consequence of the latter (the first possibility in 1), but the converse implication does not hold.\footnote{It is remarkable that the connection between the Kelvin-Planck formulation of the second law and Carath\'{e}dory's i-point principle was clarified only in the 1960s (independently) by Crawford and Oppenheim (1961) and Landsberg (1964); see also Dunning-Davies (1965).} 

In section 9 of Carath\'{e}odory's paper, the argument presupposes that adiabatic accessibility is a transitive relation between states. Since it is obviously reflexive, this implies that the relation is a \textit{preorder}. The qualification mentioned earlier in relation to section 9 is that, as originally noted by Bernstein\footnote{Bernstein (1960). See also Uffink (2001), pp. 367-368.},  the argument is of a local, not global nature in the state space; indeed entropy itself is a local notion in Carath\'{e}odory's approach.\footnote{This is related to the local nature of Frobenius' theorem. See Frankel (2004), pp. 183, 184.} Hence the adiabatic accessibility for Carath\'{e}odory is locally, not globally, a preorder. With that caveat in mind, we need to digress briefly and discuss the arrow of time in thermodynamics in order to fully understand the nature of the ambiguity in Carath\'{e}odory's system.

\section{The arrow of time}

In physical theories generally, time plays a multi-faceted role. There is the notion of temporal duration between events occurring at the same place (temporal metric, related to the ticking of an ideal inertial clock\footnote{Accelerating clocks differ from inertial ones in the sense that their behaviour generally depends on their constitution.}), the comparison of occurrences of events at different places (distant simultaneity, registered by synchronised clocks) and the directionality, or arrow, of time. Thermodynamics is of course the only theory, apart from that of the weak interactions, which incorporates this last element at a fundamental level. But first a word about the first two aspects of time within thermodynamics.

It is sometimes said that thermodynamics has no clocks, in the sense that none of its fundamental laws contains derivatives with respect to time. For example, entropy is claimed never to decrease in adiabatic processes, but the theory gives no information about how quickly changes in entropy, if any, occur. Now it may be thought that a temporal metric and a privileged notion of simultaneity both lurk in the background, because thermodynamics always appeals to the mechanical notion of work. Whether this appeal to work introduces through the back door all the temporal structure of Newtonian time is a moot point. However that may be, a noteworthy feature of Carath\'{e}odory's 1909 formulation is the fact that time derivatives do appear explicitly in his paper. This occurs in section 3 of the 1909 paper, where Carath\'{e}odory defines quasi-static adiabatic processes as those in which the thermodynamic parameters change infinitely slowly.\footnote{More specifically, quasi-static adiabatic processes are defined as those in which the difference between the work done externally and the limit of the energy change defined when the derivatives of the thermodynamic parameters converge uniformly to zero is less than experimental uncertainty. It is noteworthy that Carath\'{e}odory does not define quasi-static in the general case.}

Returning now to the ambiguity in Carath\'{e}odory's treatment of the second law, it may reasonably be asked: with respect to what arrow of time is experiment supposed to resolve the ambiguity? The claim that a certain process unfolds in such and such a way in time only makes sense in physics if some independent arrow of time is acting as a reference. What primordial arrow is being appealed to in determining whether entropy increases or decreases in non-quasi-static adiabatic processes, or similarly in determining  how internal energy changes in such processes starting and ending with the same deformation coordinates?

The view that the arrow of time in thermodynamics is defined by the traditional version of the second law has been urged by Hawking:
\begin{quote}
Entropy increases with time, because we define the direction of time to be that in which entropy increases (Hawking, 1994, p. 348).
\end{quote}
Of course, such a definition would make Carath\'{e}odory's appeal to experiment pointless. Hawking may well be articulating what is a widespread view, but it is questionable.  Within the traditional approach to thermodynamics, the mere introduction of a concept like a Carnot cycle presupposes a temporal ordering as applied to the equilibrium states within the cycle, well before questions concerning the efficiency of such cycles in relation to other kinds of heat engine are raised, and hence before the second law is postulated. In Carath\'{e}odory's formulation, even the notion of adiabatic (in)accessibility is well-defined only in relation to a background temporal ordering. What is it?
Appeal to the psychological arrow of time (according to which we remember the past not the future), for example, is unattractive; it seems plausible that formation of memories would be impossible without thermodynamic irreversibility, even if there is debate about the details.\footnote{Maroney (2010) has argued that the logical operations involved in computation do not \textit{per se} determine an arrow of time. But in a rejoinder, Smith (2014) has claimed that in the brain computational processes and the formation of long-term memories in fact require the existence of certain spontaneous diffusion/equilibration processes. From the point of view of statistical mechanics, these processes correspond to local entropy increase. But from the point of view of thermodynamics, insofar as they are macroscopic they are arguably as much tied up with the irreversible Equilibrium Principle (see below) as the Second Law.}

Arguably the simplest and perhaps most elegant answer to the question at hand is not provided by the second law, but is found nonetheless within thermodynamics itself. It has occasionally been noted that a fundamental principle that underlies all thermodynamic reasoning (including the zeroth law) is this:

\bigskip

\noindent \textit{An isolated system in an arbitrary initial state within a finite fixed volume will spontaneously attain a unique state of equilibrium.}
\bigskip

\noindent This ``Equilibrium Principle'' is the entry point in thermodynamics of time asymmetry: an isolated system evolves from non-equilibrium into equilibrium, but not the reverse.\footnote{As pointed out by Uffink (2001), Planck had already in 1897 emphasised the independence of this principle from the second law. A detailed account of the Equilibrium Principle, the emphasis it has received in the literature, and its role in the foundations of thermodynamics, is found Brown and Uffink (2001), where the principle is called the Minus First Law. In Hemmo and Shenker (2012), chapter 2, it is called the Law of Approach to Equilibrium.} So it is suggested now that all references, implicit and explicit, to the temporal ordering of events in thermodynamics can naturally be understood in relation to the arrow of time defined by the process of spontaneous equilibration. In particular, it is suggested that the kind of experiment envisaged by Carath\'{e}odory in order to resolve the ambiguity in his treatment of irreversible processes would reveal a non-decreasing entropy \textit{relative to this arrow of time}.

\section{Lieb and Yngvason}

\subsection{Introduction}
In 1999, Elliott H. Lieb and Jakob Yngvason proposed a new axiomatization of thermodynamics, which owed much to the work of  Carath\'{e}odory and that of later writers such as Robin Giles.\footnote{Lieb and Yngvason (1999); a shorter more informal version of this paper is found in Lieb and Yngvason (2000).}  The central concept is again the binary relation on the state space associated with adiabatic accessibility, now designated by $\prec$, and assumed to be \textit{globally} a preorder.\footnote{Uffink has argued that the definition of adiabatic accessibility therein is subtly different from most definitions in the literature -- and in particular that of Carath\'{e}odory; see Uffink (2001), pp. 381 \textit{ff}.  We need not dwell on this point.} But an attempt is made by the authors to provide a treatment of entropy  and its essential properties based on  ``maximum principles  instead of equations among derivatives'', so that real systems where some of these derivatives fail to be well-defined at certain points pose no special problems for the theory. Another notable and unusual feature is the attempt to provide a proof of the \textit{comparison hypothesis}, normally tacitly assumed to be an essential property of a well-behaved thermodynamic system, which states that for any states $X$ and $Y$ in the state space, then either $X \prec Y$ ($Y$ is adiabatically accessible from $X$) or $Y \prec X$. The Lieb-Yngvason (L-Y) formulation of thermodynamics is a \textit{tour de force} of physical and mathematical reasoning.

It should be noted that two further incentives behind the formulation are the desire to banish the notion of heat altogether from thermodynamics, and a shift of emphasis from \textit{impossible} processes (as in traditional formulations) to \textit{possible} ones.\footnote{The first incentive strikes us as unnecessary, given that Carath\'{e}odory had shown that heat is not fundamental, and that it is perfectly well-defined given his first law and the conservation of energy, both of which feature in the L-Y approach. Their claim that no one ``has ever seen heat, nor will it ever be seen, smelled or touched'' (Lieb and Yngvason (1999), p. 6) can also be said to apply to entropy! The second incentive  also strikes us as curious, particularly because the authors themselves stress that if irreversible processes did not exist, ``it would mean that nothing is forbidden, and hence there would be no second law.'' (\textit{ibid} section G).}

\subsection{Entropy}

Our main concern lies more with energy than entropy, but a word about the L-Y treatment of the latter is in order. This treatment, remarkably, provides what is effectively a representation theorem for the preorder $\prec$ on the state space in terms of a numerical ``entropy'' function on the space. It will be recalled that adiabatic accessibility is a temporally-ordered concept, and the question arises whether and how the monotonic temporal behaviour of entropy in adiabatic processes is connected with natural constraints on this preorder, \textit{without appeal to an assumption as strong as Carath\'{e}odory's i-point principle.} Lieb and Yngvason introduce  six plausible axioms governing the $\prec$ relation holding for single and compound systems, and assuming (without proof at this stage) the comparison hypothesis, show that there exists a real-valued function $S$ on all states of all systems such that  $X \prec Y$ if and only if $S(X) \leq S(Y)$. Furthermore, $S$ has the properties of additivity and extensivity that one expects of the entropy function. ``In a sense it is amazing'', Lieb and Yngvason write, ``that much of the second law follows from certain abstract properties of the relation among states, independent of physical details (and hence of concepts such as Carnot cycles).''\footnote{Lieb and Yngvason (1999) p. 14.} It should not be overlooked, however, that the very definition of entropy in this construction requires the existence of at least one pair of states $X_0$ and $X_1$ such that  $X_0 \prec \prec X_1$, i.e. $X_0 \prec X_1$ but not the converse. Like the traditional notion of entropy, the L-Y notion is not meaningful in a world without irreversibility of some sort.\footnote{Here we part company with Lieb and Yngvason (1999), p. 24, where it is claimed that if all states are adiabatically equivalent, entropy is constant.}

The ``Entropy Principle'' is striking, and its proof is ingenious. But it is important to note that  the temporal monotonicity associated with this numerical representation of $\prec$ does not resolve the kind of ambiguity found in Carath\'{e}odory's system. The question, recall, is whether the physical entropy is non-increasing or non-decreasing relative to the arrow of time determined by the Equilibrium Principle.  In the L-Y formulation, a formal definition of $S$ is constructed such that, given all the assumptions, $S$ cannot decrease in adiabatic processes. But the representation theorem of course holds just as well for the function $\tilde{S} \equiv - S$, in which case $X \prec Y$ if and only if $\tilde{S}(X) \geq \tilde{S}(Y)$. There is nothing in the theorem \textit{per se} which distinguishes between $S$ and $\tilde{S}$ in terms of physical import. 

\subsection{The ambiguity again}

The L-Y framework takes on a different, more geometrical tone after the treatment of the Entropy principle. The systematic treatment of irreversibility in simple systems requires additional axioms in order to derive an analogue of Carath\'{e}odory's i-point principle and notably the \textit{global} foliation of the state space defined by adiabats. 
Recall now the energy ambiguity in Carath\'{e}odory's theory outlined in section 3 above for simple systems. Precisely this issue is addressed in section III.C of the L-Y 1999 paper. The authors first adopt the view that within their framework of axioms, it is ``conventional'' whether in an adiabatic process between states with the same deformation (work) coordinates the internal energy never decreases, or never increases.\footnote{Lieb and Yngvason (1999) p. 44. There are two notions of universality involved here, and Lieb and Yngvason are anticipating two theorems. The first is Theorem 3.3, which states that the forward sectors \textit{of all states} in the state space of a simple system (i.e. the set of states adiabatically accessible from the given state) point the same way energy-wise. The second is Theorem 4.2, which states that the forward sectors \textit{of all simple systems} point the same way. It is worth emphasising that the proofs of these theorems depend on several axioms additional to the set A1-A6 needed for the Entropy Principle. The content of one of these additional axioms, A7, is discussed in the next section.} (Unsurprisingly, they adopt the former convention.) This is a curious stance, difficult to reconcile with subsequent remarks which take into account the definition of adiabatic accessibility peculiar to Lieb and Yngvason:

\begin{quote} 
From a physical point of view there is more at stake, however. In fact, our operational interpretation of adiabatic processes involves either the raising or lowering of a weight in a gravitational field and these two cases are physically distinct. Our convention, together with the usual convention for the sign of energy for mechanical systems and energy conservation, means that we are concerned with a world where adiabatic process at fixed work coordinate can never result in the raising of a weight, only in the lowering of a weight. The opposite possibility differs from the former in a mathematically trivial way, namely by an overall sign of the energy, but given the physical interpretation of the energy direction in terms of raising and lowering of weights, such a world would be different from the one we are used to.\footnote{Lieb and Yngvason (1999), p. 44.}
\end{quote}
This seems to be an admission that, as Carath\'{e}odory claimed (admittedly in the context of entropy not energy), two distinct physical possibilities are at stake, and that the issue is not merely one of convention, in the usual sense of the term. It is noteworthy that Lieb and Yngvason state as a \textit{theorem}, which they call \textit{Planck's principle}, that: 
\begin{quote} 
If two states, $X$ and $Y$, of a simple system have the same work coordinates, then $X \prec Y$ if and only if the energy of $Y$ is no less than the energy of $X$.\footnote{This is theorem 3.4, \textit{op cit}, p. 45.}
\end{quote}
The authors make a point of saying that this principle (or rather a consequence of it), which is explicitly based on their energy ``convention'', is ``clearly stronger than Carath\'{e}odory's principle, for it explicitly identifies states that are arbitrarily close to a given state, but not adiabatically accessible from it.''\footnote{\textit{op cit} p. 46.}  But such extra strength cannot be the result of adopting a convention. Note that Lieb and Yngvason give as the reason for calling the mentioned theorem ``Planck's principle'' that ``Planck emphasized the importance for thermodynamics of the fact that 'rubbing' (i.e., increasing the energy at fixed work coordinate) is an irreversible process''. Indeed the fact that frictional rubbing under fixed deformation coordinates leads to an increase of internal energy is sometimes chosen as the extra empirical ingredient needed to resolve the ambiguity in Carath\'{e}odory's original theory.\footnote{See for example Turner (1962) and Sears (1963).} 

It may be worth emphasising then the somewhat misleading nature of the claim Lieb and Yngvason make that 
Planck's principle, and as a consequence the standard Kelvin-Planck version of the second law, follow from their first nine axioms.\footnote{Lieb and Yngvason (1999) p. 46.} (These include the convex combination axiom A7 to which we return below.) This is true only when the above positive energy ``convention" is adopted. The next section attempts to further clarify the role of this extra \textit{factual} ingredient in recovering variants of the standard second law of thermodynamics within the L-Y scheme.

\subsection{The flow of energy (heat)}

If the question is whether there is any component of the L-Y scheme that adiabatically favours one side of the adiabats from the other, then the answer is actually yes. A key assumption in the L-Y treatment of simple systems and their irreversible behaviour is their 
 convex combination axiom A7. This asserts that for any two states $X_1=(U_1,V_1)$ and $X_2=(U_2,V_2)$ of a simple system ($U$ being the internal energy), a fraction $t$ of $X_1$ can be adiabatically combined with a fraction $(1-t)$ of $X_2$ to form a new state $Y=(tU_1+(1-t)U_2,tV_1+(1-t)V_2)$.\footnote{For many systems the axiom is a direct consequence of the Equilibrium Principle which implies that for any two samples of a gas in states $X_1$ and $X_2$, when you partition the samples as just described and then connect the two partitioned regions, the resulting system will eventually reach an equilibrium state $Y$. This process can be done adiabatically, and by conservation of energy state $Y$ will have energy $U=tU_1+(1-t)U_2$. By assumption, the state also has volume $V=tV_1+(1-t)V_2$. Compare with p. 31 in \textit{op. cit.}.} 
This axiom leads immediately to the theorem that the in the case of a single simple system the set of points in the state space adiabatically accessible from a given point $X$ -- the ``forward sector'' associated with $X$ -- must be a convex set.\footnote{From section III onwards in \textit{op. cit.}, the state space for a simple system is taken to be a convex subset of $\mathbf{R}^{(n+1)}$, $n$ being the number of deformation coordinates.} That is, if $X_1$ and $X_2$ as just defined are in the set, then $Y$ is also in the set. (Indeed, this is the main consequence of A7, although A7 is needed to derive several of the key geometric properties of the forward sectors.) 
Consider then a continuous curve in a $U,V$ diagram corresponding to the boundary of the set of states adiabatically accessible from a given state, which as expected in the L-Y scheme turns out to be a curve of constant entropy (see Figure 1 below).\footnote{For a discussion of the continuity of such an adiabat, see \textit{op. cit.} p. 42. Figure 1 is a special case of Figure 3 in \textit{op. cit.}, pp. 32 and 96.}
\begin{figure}[h]
\begin{center}
      \includegraphics[scale=0.25]{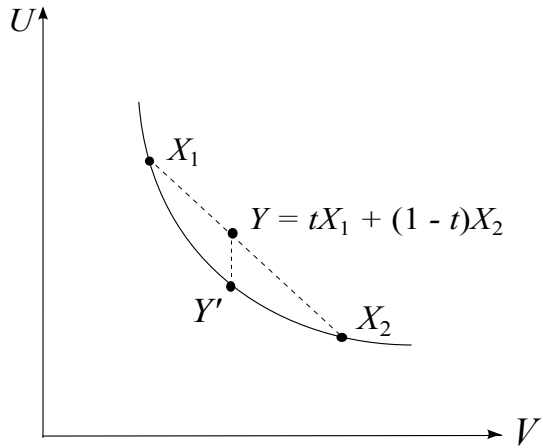}
  \caption{Constant entropy curve}
  \end{center}
\end{figure}
For a standard thermodynamic system with positive pressure, $U$ decreases on this curve with increasing deformation coordinate $V$. The dotted line between states $X_1$ and $X_2$ represents the locus of all convex combinations of these states obtained by ranging over the parameter $t $ ($0 \leq t \leq 1$). Both $X_1$ and $X_2$ are adiabatically accessible from $Y'$, and it follows from axiom A7 that $Y$ is too. In the figure, the adiabat is represented as a convex function, so $(\partial^2 U/\partial V^2)_S\geq 0$. So the forward sector defined relative to any state $X$ on the adiabatic is ``upward pointing'' in Lieb and Yngvason's terms: the projection on the energy axis of the normal to the tangent plane at $X$ pointing to the interior of the forward section is positive.\footnote{Orthogonality is defined with respect to the canonical scalar product on $\mathbf{R}^{(n+1)}$.} This is a necessary condition for the Planck principle to hold. (Note in the figure that for the state $Y$  there is a state $Y'$ on the adiabat with the same deformation coordinates, and $U(Y) \geq U(Y')$.) However, if the adiabat is concave, then the forward sector will be downward pointing. So apart from the special case of a flat boundary, owing to axiom A7 the shape of the adiabatic will determine where the forward sector lies unambiguously.\footnote{The role of A7 in determining the directionality of processes in the L-Y approach is also pointed out and explained in Henderson (2014), though in a different way to our own.}
 
However, it seems that the shape of the adiabat, and hence the upward or downward pointing nature of the forward sector, are not  determined by the L-Y axioms. Such independence indeed seems to underlie the view of Lieb and Yngvason that it is a ``convention''. 

Nor does  the convex combination axiom A7 seem to be required to obtain important variants of the second law.
Consider the simple derivation Lieb and Yngvason give of the proposition that energy (heat) spontaneously flows from hot to cold bodies, and not the converse (recall the related ambiguity in section 3 above).\footnote{See theorem 5.4, Lieb and Yngvason (1999), pp. 66-67.} A body $A$ is defined to be hotter than another body $B$ if the absolute temperature $T_A$ is greater than $T_B$, where temperature $T_{A(B)}$ is defined as $\left ( \partial S_{A(B)} / \partial U_{A(B)} \right )^{-1}$, $S_{A(B)}$ being the entropy of $A(B)$. Note that this definition of ``hotter than" is reasonable only if $T_{A(B)}$ is everywhere positive, and this constraint is a consequence\footnote{See \textit{op.cit .}pp. 44 and 62.} of the upward pointing nature of the forward sectors and the choice of the function $S$ and not $\tilde{S} \equiv - S$ in defining temperature in this way (see the end of section 5.2 above). It is easy now for Lieb and Yngvason to obtain the desired irreversible flow of energy (heat) from $A$ to $B$, given the conservation of total energy for the joint system, and the monotonicity of $U_{A(B)}$ with respect to $T_{A(B)}$. We won't repeat the details, other than to make two remarks.
 
 First, note that the spontaneity of this energy flow process (once thermal contact is achieved) leading to a common temperature is secured by appeal to a special case of the Equilibrium Principle.\footnote{Compare the discussion of the axiom $T_1$ of thermal contact in \textit{op. cit.}, p. 52.} Second, and more to the point, the strict monotonicity condition between $U_{A(B)}$ and $T_{A(B)}$ is a consequence of the concavity and differentiability of the entropy for simple systems. Concavity of entropy in general terms is established by Lieb and Yngvason's theorem 2.8, which relies on the convex combination axiom A7. However, what is actually required in the heat flow argument is the weaker claim that entropy is concave relative to the internal energy, and Lieb and Yngvason had earlier established that this does not depend on axiom A7.\footnote{\textit{Op. cit.}, p. 53.} It seems then that this axiom is not crucial to the argument, whereas the upwards pointing condition certainly is. 
 
 Axiom A7 may however give us further insight into what an anti-Kelvin world would be like in the L-Y scheme. Suppose then that the adiabat is strictly concave (not convex as depicted in Figure 1), the Planck principle is false, and $(\partial^2 U/\partial V^2)_S < 0$. Now pressure $P$ satisfies the equation  $P=-(\partial U/\partial V)_S$.\footnote{This equation is a consequence of the definition of pressure and a number of continuity assumptions in the L-Y formulation; see pp. 40-41 in \textit{op. cit.}. Note that the minus sign is omitted in equation (3.6) therein; this is not the case in Lieb and Yngvason (1998): see equation (14).}  Thus $(\partial P/\partial V)_S=-(\partial^2 U/\partial V^2) > 0$. Suppose a weight is placed on top of a piston with the pressure exactly set to balance the weight. Any bump involving the least increase of volume would increase the pressure and lead to a runaway process propelling the weight upwards; conversely a perturbation decreasing the volume would cause a continual collapse of the piston to zero volume. By Le Chatelier's Principle, this means that all the states of the system are unstable equilibria, analogous to points on the upwards-sloping region of a van der Waals isotherm.\footnote{See Reif (1965).} This instability argument also holds for negative-pressure systems with $U$ as an \emph{increasing} function of $V$ on the adiabats.
  
\section{Conclusions}

We have argued that to understand aspects of Carath\'{e}odory's, or indeed any approach to the second law in thermodynamics, a background arrow of time needs to be specified, and we suggest defining it by way of the Equilibrium Principle (Minus First Law). The L-Y approach, compared to Carath\'{e}odory's, requires weaker assumptions in order to derive a monotonic entropy function, but needs considerably more axioms in order to establish a recognisable version of the second law. In part this reflects an admirable attention to detail and an attempt to make every step transparent while using less differential structure; one must also not overlook the ambitious program of deriving the comparison hypothesis. However, a strict analogue of the energy ambiguity in Carath\'{e}odory's approach reappears in the L-Y scheme. The upward pointing nature of forward sectors (and hence the Planck principle) appears not to be a consequence of the L-Y axioms, and nor is it a mere convention. It plays the role of an appeal to experience, over and above the axioms, of the kind Carath\'{e}odory needed in order to derive the standard Kelvin-Planck version of the second law.

\section{Acknowledgments}

We thank Leah Henderson, Elliott Lieb, Jos Uffink, and Jakob Yngvason for helpful discussions.

-----------------------------------------

B. Bernstein (1960), ``Proof of Carath\'{e}odory's Local Theorem and its Global
Application to ThermostaticsÕ, \textit{Journal of Mathematical Physics} \textbf{1}, 222-224.

J. B. Boyling (1972), ``An axiomatic approach to classical thermodynamics'', \textit{Proceedings of the Royal Society of London A}, \textbf{329}, 35-70.

Harvey R. Brown and Jos Uffink (2001), ``The Origins of Time Asymmetry in Thermodynamics: The Minus First Law'', \textit{Studies in History and  Philosophy of Modern Physics} \textbf{32} No. 4, pp. 525-538.

Constantin Carath\'{e}dory (1909), ``Untersuchungen \"{u}ber die Grundlagen der Thermodynamik'', \textit{Mathematische Annalen} (Berlin) \textbf{67}, 355-386. English translation by Joseph Kestin: ``Investigation into the Foundations of Thermodynamics'', in Joseph Kestin (ed.), \textit{The Second Law of Thermodynamics: Benchmark papers on energy, Vol. 5}, Dowden, Hutchinson and Ross, Stroudsberg Pennsylvania (1976), pp. 229-256. Page numbers refer to the translation.

Constantin Carath\'{e}dory (1925), ``\"{U}ber die Bestimmung der Energie und der absoluten
Temperatur mit Hilfe von reversiblen Prozessen'', \textit{Sitzungsberichte der Preussischen
Akademie der Wissenschaften, Physikalisch-Mathematische Klasse 1}, pp. 39-47.

B. Crawford, Jr. and I. Oppenheim, ``The Second Law of Thermodynamics'', \textit{The Journal of Chemical Physics}, \textbf{34} (5), 1621-1623.

J. Dunning-Davies (1965), ``Carath\'{e}odory's Principle and the Kelvin Statement of the Second Law'', \textit{Nature}, \textbf{208}, 576-577. doi:10.1038/208576a0

Theodore Frankel (2004), \textit{The Geometry of Physics}, Cambridge University Press, New York; second edition.

Meir Hemmo and Orly R. Shenker (2012), \textit{The Road to Maxwell's Demon: Conceptual Foundations of Statistical Mechanics}; Cambridge University Press, Cambridge UK.

Leah Henderson (2014), ``Can the second law be compatible with time reversal invariant dynamics?'', to appear in \textit{Studies in History and Philosophy of Modern Physics}.

Peter T. Landsberg (1956), ``Foundations of Thermodynamics", \textit{Reviews of Modern Physics} \textbf{28}, 363-392. DOI: http://dx.doi.org/10.1103/RevModPhys.28.363

Peter T. Landsberg (1961), ``On suggested simplifications of Carath\'{e}odory's thermodynamics", \textit{physica status solidi (b)} \textbf{1} (2), 120-126. 10.1002/pssb.19610010203.

Peter T. Landsberg (1964), ``A Deduction of CARATH\'{E}ODORY'S Principle from Kelvin's Principle", \textit{Nature} \textbf{201}, 485-486. 

Peter T. Landsberg (1990), \textit{Thermodynamics and Statistical Mechanics}; Dover Publications, Inc., New York.

Elliott H. Lieb and Jakob Yngvason (1998), ``A guide to entropy and the second law of thermodynamics'',  arXiv:math-ph/9805005v1. 

Elliott H. Lieb and Jakob Yngvason (1999), ``The physics and mathematics of the second law of thermodynamics'', \textit{Physics Reports} \textbf{310}, 1-96. The 1999  online version of this paper, to which our page numbers refer, is at arXiv:cond-mat/9708200v2.

Elliott H. Lieb and Jakob Yngvason (2000), ``A fresh look at entropy and the second law of thermodynamics'', \textit{Physics Today} \textbf{53}, 32-37. 

Owen J. E. Maroney (2010), ``Does a Computer Have an Arrow of Time?'', \textit{Foundations of Physics} \textbf{40}, 205-238.

Frederick Reif, (1965), \emph{Fundamentals of Statistical and Thermal Physics}, McGraw-Hill, Inc., Boston MA..

A. A. Robb (1921), \textit{The Absolute Relations of Time and Space}, Cambridge University Press.

Arthur E. Ruark (1925), ''LXIII. The proof of the corollary of carnot's theorem'', \textit{Philosophical Magazine Series 6}, \textbf{49:291}, 584-585. 10.1080/14786442508634639.

Francis W. Sears (1963), ``A Simplified Simplification of Carath\'{e}odory's Treatment of Thermodynamics'', \textit{American Journal of Physics}, \textbf{31} (10), 747-752.

Francis W. Sears (1966), ``Modified Form of Carath\'{e}odory's Second Axiom'', \textit{American Journal of Physics}, \textbf{34}, 665-666.

Ryan Smith (2014) ``Do Brains Have an Arrow of Time?", \textit{Philosophy of Science} \textbf{81}, 265-275. 0031-8248/2014/8102-0003.

Louis A. Turner (1960) ``Simplification of Carath\'{e}odory's Treatment of Thermodynamics'', \textit{American Journal of Physics}, \textbf{28}, 781-786.

Louis A. Turner (1962) ``Simplification of Carath\'{e}odory's Treatment of Thermodynamics. II'', \textit{American Journal of Physics}, \textbf{30}, 506-508.

Louis A. Turner (1963) ``Temperature and Carath\'{e}odory's Treatment of Thermodynamics'', \textit{The Journal of Chemical Physics} \textbf{38}, 1163-1167. 10.1063/1.1733818

Jos Uffink (2001), ``Bluff Your Way in the Second Law of Thermodynamics'', \textit{Studies in
History and Philosophy of Modern Physics} \textbf{32} (3), 305-394.

Jos Uffink (2003), ``Irreversibility and the Second Law of Thermodynamics'', in \textit{Entropy}, Andreas Greven, Gerhard Keller, and Gerald Warnecke (eds.), Princeton Series in Applied Mathematics, Princeton University Press, New Jersey; pp. 121-146.

\end{document}